\documentclass[aps,pra,preprint,showpacs,showkeys,nofootinbib]{revtex4}
\usepackage{amsfonts}
\usepackage{amsmath}
\usepackage{amssymb}
\usepackage{graphicx}%
\usepackage{epsfig}
\providecommand{\U}[1]{\protect\rule{.1in}{.1in}}

\newcommand{\rmd}{\mathrm{d}}
\newcommand{\rmi}{\mathrm{i}}
\newcommand{\rme}{\mathrm{e}}

\begin{document}
\preprint{ }
\title{Tailoring high-order harmonic generation with nonhomogeneous fields and electron confinement}
\author{M. F. Ciappina$^{1}$}
\author{Srdjan S. A\'{c}imovi\'{c}$^{1}$}
\author{T. Shaaran$^{1}$}
\author{J. Biegert$^{1,2}$}
\author{R. Quidant$^{1,2}$}
\author{M. Lewenstein$^{1,2}$}
\affiliation{$^{1}$ICFO-Institut de Ci\`ences Fot\`oniques, 08860 Castelldefels (Barcelona), Spain}
\affiliation{$^{2}$ICREA-Instituci\'o Catalana de Recerca i Estudis Avan\c{c}ats, Lluis Companys 23, 08010 Barcelona, Spain}

\keywords{high-order harmonics generation;strong field approximation; nanostructures; plasmonics}
\pacs{42.65.Ky,78.67.Bf, 32.80.Rm}
\begin{abstract}
We study high-order harmonic generation (HHG) resulting from the illumination of plasmonic nanostructures with a short laser pulse. We show that both the inhomogeneities of the local electric field and the confinement of the electron motion play an important role in the HHG process and lead to a significant increase of the harmonic cutoff. In order to understand and characterize this feature, we combine the numerical solution of the time dependent Schr\"odinger equation (TDSE) with the electric fields obtained from 3D finite element simulations. We employ time-frequency analysis to extract more detailed information from the TDSE results and to explain the extended harmonic spectra. Our findings have the potential to boost up the utilization of HHG as coherent extreme ultraviolet (XUV) sources. 
\end{abstract}
\maketitle

When atoms and molecules are subject to intense laser radiation, new phenomena appear as a consequence of this interaction. Among them, high-harmonic
generation (HHG), above threshold ionization (ATI), and non-sequential double ionization (NSDI) can
be mentioned as the most important ones~\cite{keitel,krausz}. In particular HHG represents the most reliable pathway to coherent light sources in the ultraviolet (UV) to extreme ultraviolet (XUV) spectral range. These tools are in high demand nowadays for basic research, material science, biology and possibly lithography~\cite{misharmp}. Their principal complication is the demanding infrastructure needed for XUV generation and target delivery as well as its low efficiency and low duty cycle. The recent demonstration based on surface plasmon resonances as light enhancers could provide a potential solution to this problem~\cite{kim}. 

The physical mechanism behind the generation of high-order harmonics has been well established in the so-called three step or simple man's model~\cite{corkum,sfa}: The first step is the strong-field ionization of the atom or molecule as a consequence of the nonperturbative interaction with the coherent electromagnetic radiation. The classical propagation of the electron in the field
establishes the second step of this original model. Lastly, the third step in the progression occurs when the electron is steered back in the
linearly polarized field to its origin, recombining under the emission of a high-energy photon. One of the main features
of the HHG process is the coherence of the emitted radiation, which, e.g., opens the possibility of generating attosecond
pulses~\cite{corkumnat} or to extract temporal and spatial information with attosecond and sub-Angstrom resolution, respectively~\cite{manfred_rev}.

Field enhanced HHG using plasmonics, generated starting from engineered metal nanostructures, requires no extra cavities or laser pumping to amplify the power of the input pulse. By exploiting surface plasmon resonances, local electric fields can be enhanced by more than 20 dB~\cite{muhl,schuck}. Consequently, the intensity of the enhanced local electric field is strong enough to exceed the threshold laser intensity for HHG generation in noble gases. In particular, using gold bow-tie shaped nanostructures, it is shown that the enhancement is sufficient to produce, starting with a laser source of 800 nm, XUV wavelengths from the 7th (114 nm) to the 21st (38 nm) harmonics and the pulse repetition rate remains unaltered without any extra pumping or cavity attachment. Furthermore, the high harmonics radiation generated from each nanostructure acts as a point-like source, enabling collimation or focusing of this coherent radiation by means of (constructive) interference. This opens a wide range of possibilities to spatially arrange nanostructures to enhance or shape spectral and spatial properties in numerous ways~\cite{kim}.

The basic principle of HHG based on plasmonics can be outlined as follows (the full explanation can be found in~\cite{kim}). A femtosecond low intensity laser pulse is coupled to the plasmon mode inducing a collective oscillation of free charges within the metal. The free charges redistribute the electric
field around each of the metal nanostructure, thereby forming a spot of highly enhanced electric field. The enhanced field exceeds largely the
threshold of HHG, thus by injection of noble gases onto the spot of the enhanced field, high
harmonics are generated. In here the enhanced field is not spatially homogeneous in the region the electron dynamics will take place. Additionally the spatial region where the electron moves is restricted in space. These two features imply strong modifications in the harmonic spectra, as was shown recently by several authors~\cite{husakou,yavuz,ciappi2012}.

Up to now numerical and semiclassical approaches to study laser-matter processes in atoms and molecules, in particular high-order harmonic generation (HHG), are largely based on the dipole approximation in which the laser electric field ($\mathbf{E}(\mathbf{r},t)$) and its vector potential associated ($\mathbf{A}(\mathbf{r},t)$) are spatially homogeneous in the region where the electron dynamics takes place, i.e. $\mathbf{E}(\mathbf{r},t)=\mathbf{E}(t)$ and $\mathbf{A}(\mathbf{r},t)=\mathbf{A}(t)$~\cite{keitel,krausz}. On the other hand, the fields generated using plasmonics are spatially dependent and can not be described by the dipole approximation. From a theoretical viewpoint, the HHG process using homogeneous fields can be tackled using different approaches (for a summary see e.g.~\cite{book1,book2} and references therein). In this Letter we extend the Time Dependent Schr\"odinger Equation (TDSE) in order to study the harmonic radiation generated by a model atom when it is illuminated by a spatially inhomogeneous electric field. In order to generate this field we consider metal bow-tie shaped nanostructures as those used in~\cite{kim}. For a linearly polarized field, which is the case of our study, the dynamics of an atomic electron is mainly along the direction of the field and as a result it is a good approximation to employ the Schr\"odinger equation in one spatial dimension (1D-TDSE)~\cite{keitel} which reads:
\begin{eqnarray}
\label{tdse}
\rmi \frac{\partial \Psi(x,t)}{\partial t}&=&\mathcal{H}(t)\Psi(x,t) \\
&=&\left[-\frac{1}{2}\frac{\partial^{2}}{\partial x^{2}}+V_{atom}(x)+V_{laser}(x,t)\right]\Psi(x,t) \nonumber
\end{eqnarray}
where $V_{atom} (x)$ is the atomic potential and $V_{laser}(x,t)$ represents the potential due to the laser electric field.
In here, we use for $V_{atom}$ the quasi-Coulomb potential
\begin{eqnarray}
\label{atom}
V_{atom}(x)&=&-\frac{1}{\sqrt{x^2+\xi^2}}
\end{eqnarray}
which first was introduced in~\cite{eberly} and has been widely used in the 1D studies of laser-matter processes in atoms. The required ionization potential can be defined varying the parameter $\xi$ in Eq. (\ref{atom}). The
potential $V_{laser}(x,t)$ due to the laser electric field $E(x,t)$ is given by 
\begin{eqnarray}
\label{vlaser}
V_{laser}(x,t)&=&E(x,t)\,x
\end{eqnarray}
with 
\begin{equation}
\label{electric}
E(x,t)=E_0\,f(t)\, h(x)\,\sin\omega t,
\end{equation}
which is linearly polarized along the $x$-axis. In Eq. (\ref{electric}), $E_0$, $\omega$ and $f(t)$ are the peak amplitude, the frequency of the coherent electromagnetic radiation and the pulse envelope, respectively. In here, $h(x)$ represents the functional form of the nonhomogeneous electric field and it can be written as a series of the form $h(x)=\sum_{i=0}^{N}b_i x^{i}$. The coefficients $b_i$ are obtained by fitting the real electric field that results from a finite element simulation considering the real geometry of different nanostructures. In this work we use for the laser pulse a trapezoidal envelope given by
\begin{equation}
\label{ft}
  f(t) = \left\{
  \begin{array}{l l}
    \frac{t}{t_{1}} & \quad \text{for $0 \leq t < t_1$}\\
    1 & \quad \text{for $t_1 \leq t \leq t_2$}\\
    -\frac{(t-t_3)}{(t_{3}-t_{2})} & \quad \text{for $t_2 < t \leq t_3$}\\
    0 & \quad \text{elsewhere}\\
  \end{array} \right.
\end{equation}
where $t_1=2\pi n_{on}/\omega$, $t_2=t_1+2\pi n_{p}/\omega$, and $t_3=t_2+2\pi n_{off}/\omega$. $n_{on}$, $n_p$ and $n_{off}$ are the number of cycles of turn on, plateau and turn off, respectively.


We use $\xi=1.18$ in Eq. (\ref{atom}) such that the binding energy of the ground state of the 1D Hamiltonian
coincides with the (negative) ionization potential of Ar, i.e. $\mathcal{E}_{GS} = -15.7596$ eV ($-0.58$ a.u.). Furthermore we assume that the noble gas atom is in its initial state (ground state (GS)) before we turning the laser ($t=-\infty$) on. Equation (\ref{tdse}) is solved numerically by using the Crank-Nicolson scheme~\cite{keitel}. In addition, to avoid spurious reflections from the spatial boundaries, at each time step, the electron wave function is multiplied by a mask function~\cite{mask}. 

The harmonic yield of an atom is proportional to the Fourier transform of the acceleration $a(t)$ of its active electron~\cite{schafer}. That is,
\begin{equation}
D(\omega)=\left| \frac{1}{\tau}\frac{1}{\omega^2}\int_{-\infty}^{\infty}\rmd t\rme^{-\rmi \omega t}a(t)\right|^2
\end{equation}
with $a(t)$ is obtained by using the following commutator relation 
\begin{equation}
\label{accel1D}
a(t)=\frac{\rmd^{2}\langle x \rangle}{\rmd t^2}=-\langle \Psi(t) | \left[ \mathcal{H}(t),\left[ \mathcal{H}(t),x\right]\right] | \Psi(t) \rangle.
\end{equation}
In here, $\mathcal{H}(t)$ and $\Psi(x,t)$ are the Hamiltonian and the electron wave function defined in Eq. (\ref{tdse}), respectively. The function $D(\omega)$
is called the dipole spectrum, which gives the spectral profile measured in HHG experiments. For solving Eq. (\ref{tdse}), the gap size $g$ of the gold bow-tie nanostructure is taken into account restricting the spatial grid size (see Figure 1 for a sketch of the gold bow-tie nanostructure including the typical dimensions and the geometry). 



The electric field intensity distribution inside the gap of the gold bow-tie nanoantenna was computed numerically by 3D Finite Element Method (COMSOL Multiphysics)~\cite{srdjan1}, using the gold optical properties taken from Ref.~\cite{christy} . The antenna is formed by two identical (isosceles) triangular gold pads (longest altitude of 600 nm and the smallest acute angle of 30$^{\circ}$) separated by an air gap $g$ (as shown in Figure 1). The apices at corners were rounded (10 nm radius of curvature) to account for limitation of current fabrication techniques and avoid nonphysical fields enhancement due to tip-effect. The out of plane thickness is set to 25 nm. These parameters yield to a dipolar bonding resonance centered at around $\lambda=1800$ nm when considering gaps ranging between 12 nm and 15 nm. This particular value of $\lambda$ was chosen according to the availability of laser sources~\cite{jens1}. On the other hand, the selected laser wavelength allows the electron to have excursions of the order of the gap $g$ and consequently to \textit{confine} its motion. Classically the electron excursion in an oscillating electric field is given by the so-called quiver radius $\alpha_0$, which is $\propto \sqrt{I}\lambda^2$ where $I$ is the laser intensity. For instance, for intensities $I$ of $\sim 10^{14}$ W cm$^{-2}$, $\alpha_0$ can have a value about $\pm 80$ a.u. ($\pm$4.5 nm).

The insets of Figs. 2 and 3 display the calculated electric field intensity enhancement in the gap of the bow-tie structures when illuminated by a linearly polarized ($x$-axis) plane wave at 1800 nm. The field-enhancement profile is extracted for the bow-tie long axis through the middle of the gap, so the successive problem is reduced to 1D. Additionally we normalize the electric field by setting $E(0,t)=1$. We observe a typical amplification between 30-40 dB, i. e. 3-4 orders of magnitude between the input intensity and the intensity at center of the gap. In a real experiment, however, the enhancement can be smaller than our calculations, nevertheless, for a similar system, it was shown that one can obtain values of more than 20 dB~\cite{kim}. 

Figures 2 and 3 depict the harmonic spectra for bow-ties shaped nanostructures with gaps $g=12$ nm and $g=15$ nm, respectively and for a laser wavelength of $\lambda=1800$ nm considering an homogeneous electric field, i.e $E(x,t)=E(t)$ and a nonhomogeneous electric field using Eq. (\ref{electric}). The laser intensities are $I=8\times 10^{13}$ W cm$^{-2}$ and $I=1.25\times 10^{14}$ W cm$^{-2}$ at the center of the spot ($x=0$), respectively. In order to reach these values, we consider enhancements between 25 and 35 dB with input intensities in the range of $2.5\times10^{11}$-$2.5\times10^{10}$ W cm$^{-2}$ for the first case and $4\times10^{11}$-$4\times10^{10}$ W cm$^{-2}$ for the second one. These intensities would be well below the damage threshold of the nanostructure employed (see e.g.~\cite{kim}).  In both cases, we use a trapezoidal shaped pulse with three optical cycles turn on ($n_{on}=3$) and turn off ($n_{off}=3$) and a plateau with 4 optical cycles ($n_{p}=4$), i.e. 10 optical cycles in total which is about 60 fs. 

For the homogeneous case we have an harmonic cutoff at around 139$\omega$ and 204$\omega$ as shown by arrows in Figs. 2 and 3, respectively. In fact, our calculation are in excellent agreement with the semiclassical model~\cite{sfa}. For nonhomogeneous cases, however, we observe a substantial increase in the harmonic cutoff,  which is about 50 \% higher than the cutoff generated by a homogeneous electric field. This new feature emerges due to the combination of the nonhomogeneous character of the electric field and the confinement of the electron motion~\cite{ciappi2012}.   


In the following, we employ time-analysis and classical calculations in order to investigate the harmonic spectra shown in Figs. 2 and 3. The results of time-analysis are presented in this Letter while the classical model will be part of the Supplemental material. To perform the former case, we employ the Gabor transformation which was developed in the 1940s by D. Gabor~\cite{gabor}. It has been proven that this technique is appropriate to estimate the emission times of harmonic spectra in atoms and molecules and to discriminate the different electron trajectories~\cite{manfred}. Starting from the dipole acceleration $a(t)$ of Eq. (\ref{accel1D}), the Gabor transform is defined as
\begin{eqnarray}
a_{G}(\Omega,t)&=&\int dt' a(t') \frac{\exp\left[-(t-t')^{2}/2\sigma^{2} \right]}{\sigma \sqrt{2 \pi}}\exp(i \Omega t')
\end{eqnarray}
where the integration is usually taken over the pulse duration. In our studies we use $\sigma=1/3 \omega$, with $\omega$ being the central laser frequency. The chosen value of $\sigma$ allows us to achieve an adequately balance between the time and frequency resolutions  (see Ref.~\cite{manfred} for details). In Figure 4 we display the Gabor analysis of the harmonic spectra of Fig. 2 and 3. Panel (a) and (c) represent the homogeneous cases corresponding to Figs. 2 and 3, respectively, while panels (b) and (d) show their nonhomogeneous counterparts.


As a well known fact, in the high-order harmonic generation both short and long electron trajectories contribute to the harmonic spectra~\cite{manfred}. In here, however, we observe that only the short electron trajectories are present as shown in Fig. 4. The absence of the long trajectories is a consequence of the electron motion in the confined region formed by the bow-tie nanostructure. In addition, for the nonhomogeneous cases, we observe an extension of the harmonic cutoff as shown in panels (b) and (d). On the other hand, our calculations show that without confining the electron motion the harmonic cutoff disappear (for more details see Supplementary material and Ref.~\cite{ciappi2012}). 

We present high-order harmonic generation of Ar produced by the fields generated when a gold bow-tie nanostructure is illuminated by a short laser pulse. The functional form of these fields is extracted from finite element simulations using both the complete geometry of the metal nanostructure and laser wavelength. We use the numerical solution of the time dependent Schr\"odinger equation (TDSE) in reduced dimensions to predict the harmonic spectra. We observe an extension in the harmonic cutoff position that could lead to the production of XUV coherent laser sources and opening the avenue to the generation of attosecond pulses. This new feature is a consequence of the combination of a nonhomogeneous electric field, which modifies substantially the electron trajectories, and the confinement of the electron dynamics. Furthermore, our numerical results are supported by time-analysis and classical simulations. A more pronounced increment in the harmonic cutoff, in addition with an appreciable growth in the conversion efficiency, could be attained optimizing the nanostructure geometry and by choosing the adequate materials.    

We acknowledge the financial support of the MICINN projects (FIS2008-00784 TOQATA, Consolider Ingenio 2010 QOIT, SAUUL CSD 2007-00013, FIS2008-06368-C02-01 and FIS2010-12834); ERC Advanced Grant QUAGATUA, Alexander von Humboldt Foundation and Hamburg Theory Prize (M. L.); LASERLAB-EUROPE (Grant 228334, EC's Seventh Framework Programme) (J. B.); ERC-2010-StG Plasmolight (Grant 259196) (R. Q.). This research has been partially supported by Fundaci\'o Privada Cellex. M.F.C. thanks T. Kirchner and M. Schulz for useful comments and suggestions.


\newpage
Figures captions \newline

Fig. 1. (Color online) (a) Schematic representation of the geometry of the considered nanostructure. A gold bow-tie antenna resides on glass
substrate (refractive index $n = 1.52$) with superstate medium
of air ($n = 1$). The characteristic dimensions of the system and the coordinate system used in the 1D-TDSE simulations
are shown. (b) Scanning Electron Microscope (SEM) image of a nanofabricated bow-tie antenna (thickness is twice larger micrograph for the increased contrast purpose)

Fig. 2. (Color online) High-order harmonic generation (HHG) spectra for Ar with ionization potential $\mathcal{E}_{GS}=-0.58$ a.u., laser wavelength $\lambda=1800$ nm and intensity $I=8\times10^{13}$ W$\cdot$cm$^{-2}$ at the center of the gap $x=0$.  We use a trapezoidal shaped pulse, Eq. (\ref{ft}), with $n_{on}=3$, $n_{off}=3$ and $n_{p}=4$ (about 60 fs). 
The gold bow-tie nanostructure has a gap $g=12$ nm (226 a.u.). Black line indicates the homogeneous case while red line indicates the nonhomogeneous case. The arrow indicates the cutoff predicted by the semiclassical model for the homogeneous case~\cite{sfa}. The top left inset shows the functional form of the electric field $E(x,t)$ where the solid lines is the raw data obtained from the finite element simulations and the dash line is a nonlinear fitting. The top right inset shows the intensity enhancement in the gap region of the gold bow-tie nanostructure.

Fig. 3. (Color online) Idem Fig. 2 but now the gold bow-tie nanostructure has a gap $g$ of 15 nm (283 a.u.) and the laser intensity is $I=1.25\times10^{14}$ W$\cdot$cm$^{-2}$ at the center of the gap $x=0$.

Fig. 4. (Color online) Gabor analysis for the harmonic spectra of Figs. 2 and 3. Panels (a) and (b) correspond to the Fig. 2 for the homogeneous and nonhomogeneous case, respectively.
While panels (c) and (d) correspond to the Fig. 3 for the homogeneous and nonhomogeneous case, respectively.  In all panels, the zoomed regions show a time interval during the laser pulse (Ref.~\cite{manfred} for details).

\end{document}